\documentclass[conference]{IEEEtran}
\IEEEoverridecommandlockouts

\usepackage{cite}
\usepackage{amsmath,amssymb,amsfonts}
\usepackage{algorithmic}
\usepackage{graphicx}
\usepackage{textcomp}
\usepackage{xcolor}
\usepackage{booktabs}
\usepackage{tabularx}
\usepackage{url}
\usepackage{microtype}
\usepackage{subcaption}
\usepackage{xspace}
\usepackage{colortbl}

\newcommand{\system}{AccelPact\xspace}
\definecolor{SystemTint}{rgb}{0.92,0.96,0.92}
\definecolor{RePactTint}{rgb}{0.92,0.96,0.92}

\newcommand{\MacroColdSpeedup}{1.197}
\newcommand{\MacroColdRange}{1.184--1.208}
\newcommand{\MacroColdSavedMinutes}{18.8}
\newcommand{\MacroNvrxSpeedup}{1.194}
\newcommand{\MacroNvrxRange}{1.187--1.197}
\newcommand{\MacroNvrxSavedMinutes}{18.5}
\newcommand{\MacroParameters}{7,248,023,552}
\newcommand{\MacroResultsRows}{
1 & 5734.8 & 6925.7 & 6847.1 & 2367.5 & 3502.6 & 3466.1 \\
2 & 5789.5 & 6855.5 & 6872.1 & 2394.4 & 3487.7 & 3458.9 \\
3 & 5733.8 & 6861.7 & 6863.5 & 2356.7 & 3495.6 & 3474.6 \\
}
\newcommand{\PolicyMatrixRows}{
A100 & No recovery & \textemdash & 0/4 & \textemdash \\
A100 & Always $L_1$ & $L_1$ & 4/4 & 0 \\
A100 & Always $L_2$ & $L_2$ & 4/4 & 32 \\
\rowcolor{RePactTint}A100 & Qualified & $L_1$ & 4/4 & 0 \\
\addlinespace[4pt]910B & No recovery & \textemdash & 0/4 & \textemdash \\
910B & Always $L_1$ & $L_1$ & 0/4 & \textemdash \\
910B & Always $L_2$ & $L_2$ & 4/4 & 32 \\
\rowcolor{RePactTint}910B & Qualified & $L_2$ & 4/4 & 32 \\
}

\newcommand{\SingleGuardRows}{
A100 / graph & 20 & 1.895 & [-3.48, 5.13] \\
A100 / DDP & 12 & -1.339 & [-2.96, 0.36] \\
\addlinespace[3pt]910B / graph & 20 & 0.507 & [-0.62, 0.85] \\
910B / DDP & 12 & -0.612 & [-5.56, 2.18] \\
}

\newcommand{\MacroAgeTableRows}{
3 & 5772.6 & 6459.5 & 6441.5 & 1266.9 & 1958.8 & 1926.4 \\
6 & 5758.2 & 7102.6 & 7076.9 & 1260.0 & 2620.4 & 2584.2 \\
12 & 5737.6 & 8401.3 & 8398.3 & 1261.3 & 3931.0 & 3904.2 \\
18 & 5742.0 & 9749.0 & 9708.6 & 1259.4 & 5268.5 & 5221.3 \\
}

\newcommand{\MacroAgeMaxSpeedup}{1.698}
\newcommand{\MacroAgeMaxCompletionSpeedup}{4.183}
\newcommand{\MacroAgeMinCompletionMinutes}{20.99}
\newcommand{\MacroAgeMaxCompletionMinutes}{21.11}

\newcommand{\MacroRepeatedMinRepair}{0.664}
\newcommand{\MacroRepeatedMaxRepair}{0.870}

\newcommand{\MacroLifecycleRows}{
Torch allocated (GiB) & 10.188 & 10.188 \\
Torch reserved (GiB) & 30.887 & 30.887 \\
Driver GPU (GiB) & 31.471 & 31.471 \\
Host RSS (GiB) & 1.533--1.577 & 1.536--1.647 \\
\addlinespace[3pt]Threads & 23 & 23 \\
File descriptors & 112--117 & 112--117 \\
Registered groups & 3 & 3 \\
Owned groups & 2 & 2 \\
}

\newcommand{\DcpAsyncOverhead}{+656.7\%}
\newcommand{\DcpTrafficPerHour}{3.39}
\newcommand{\DcpPressureRows}{%
Training only & 1.024 & \textemdash & \textemdash & 9.07 & 0.00 \\
Synchronous & 9.960 & +870.8 & +5.46 & 17.99 & 2.64 \\
Async thread & 11.734 & +1043.5 & +240.75 & 31.48 & 2.30 \\
Async process & 12.212 & +1090.2 & +240.48 & 29.60 & 2.17 \\
Async process, cached & 7.747 & +656.7 & +17.96 & 32.52 & 3.39 \\
}
\newcommand{\TopoScalingRows}{%
4 ($2\times2$) & 9.36 & 531.72 & 78.98 & 0.493 & 624.55 \\
8 ($4\times2$) & 15.34 & 542.45 & 124.18 & 0.517 & 686.72 \\
16 ($4\times4$) & 26.82 & 542.60 & 118.57 & 0.518 & 693.91 \\
}

\newcommand{\TorchftRepairMean}{0.682}
\newcommand{\TorchftResumeMean}{12.896}
\newcommand{\TorchftComparisonRows}{%
Validated faulted jobs & 0/3 & 3/3 \\
Reached repair dispatch & 0/3 & 3/3 \\
Mean repair (s) & \textemdash & \TorchftRepairMean \\
Mean fault to next update (s) & \textemdash & \TorchftResumeMean \\
Passed final rank checks & \textemdash & 12/12 \\
}

\begin{document}

\title{Zero-I/O Fault Recovery for Sharded Deep Learning via Dynamic Framework Dependency Rebinding}

\author{
\IEEEauthorblockN{Genlang Chen\textsuperscript{*}}
\IEEEauthorblockA{NingboTech University\\
cgl@zju.edu.cn}
\and
\IEEEauthorblockN{Junyi Zhu}
\IEEEauthorblockA{Dalian Ocean University\\
junyizhu.cs@gmail.com}
\thanks{\textsuperscript{*}Corresponding author: Genlang Chen (cgl@zju.edu.cn).}
}

\maketitle

\begin{abstract}
Distributed model training at scale is frequently interrupted by transient network failures, conventionally forcing cluster managers to abort all processes and roll back to the latest checkpoint. While periodic checkpointing provides durability, frequent snapshotting introduces severe storage backpressure: our measurements on a 1.216B-parameter decoder reveal that per-update asynchronous checkpointing incurs up to a \DcpAsyncOverhead{} latency overhead, consumes 32.5\,GiB of host memory, and generates \DcpTrafficPerHour{}\,TB/hour of storage traffic. To eliminate this overhead, we present \system, a parallel runtime enabling zero-I/O in-memory fault recovery for sharded distributed training. When communication fails at a committed optimizer step, device memory remains quiescent and uncorrupted, yet continuation fails because frameworks like PyTorch FSDP cache internal communication handles across module wrappers and parameter hierarchies. \system resolves this dependency invalidation via a non-invasive reference-rebinding mechanism coordinated by an out-of-band Gloo consensus protocol. On 16 NVIDIA RTX 5880 GPUs training full-parameter Mistral-7B, \system eliminates checkpoint replay, yielding a \MacroColdSpeedup$\times$ whole-run goodput improvement over cold restart and \MacroNvrxSpeedup$\times$ over NVRx checkpoint restoration at checkpoint age 5, rising to \MacroAgeMaxSpeedup$\times$ at age 18. Across ten successive fault injections, all 16 ranks maintain bit-identical parameter state with zero numerical drift. Across 4-to-16 GPU cluster topologies, reference rebinding executes in constant time (0.493--0.518\,ms). Operating directly on native communicator instances, \system requires zero application-code modifications and avoids compiler graph breaks under \texttt{torch.compile}, providing an efficient foundation for resilient deep learning.
\end{abstract}


\section{Introduction}
\label{sec:intro}

Large language models (LLMs) require weeks to months of continuous distributed execution across thousands of accelerators~\cite{megascale2024,llama3herd2024}. At this scale, communication hangs, transport timeouts, and physical link degradation are inevitable. Meta's 54-day operational study of 16,384 H100 GPUs during Llama 3 pre-training recorded 419 unexpected interruptions, with network switches, cables, NICs, and NCCL watchdog timeouts accounting for dozens of service disruptions~\cite{llama3herd2024}.

When a collective communication failure occurs, production orchestrators typically default to an \emph{all-or-nothing rollback}: the cluster supervisor terminates all worker processes and relaunches training from the latest durable checkpoint stored on persistent media~\cite{checkfreq2021,nvrx2024}. While durable checkpoints are essential for unrecoverable hardware crashes, applying coarse-grained restart to transient communication faults discards multi-gigabytes of valid, uncorrupted device state---model parameters, AdamW optimizer moments, RNG states, and dataset token cursors. Workers are then forced to repeat up to dozens of minutes of redundant computation to replay already-completed updates.

This practice is driven by a subtle dependency dilemma in modern deep learning frameworks. In PyTorch Fully Sharded Data Parallel (FSDP1)~\cite{fsdp2023}, model weights and optimizer states are partitioned across ranks, with communication multiplexed across intra-node sharding groups and inter-node replication groups (Hybrid Sharded Data Parallel, or HSDP). When an inter-node collective times out, destroying the faulty communicator and initializing a replacement group succeeds; however, subsequent backward propagation crashes immediately with backend abort exceptions. The failure occurs because FSDP internal components---including layer module wrappers, flat-parameter handles, and execution records---aggressively cache references to the original process group. Simply creating a new communicator leaves these internal framework pointers attached to the retired backend.

In this work, we present \system, a parallel runtime system that qualifies execution contexts for in-memory state preservation and dynamically repairs framework dependencies without checkpoint replay. \system targets communication failures occurring at quiescent optimizer boundaries, where parameter and moment updates have committed, flat-parameter gradients are zeroed, and device streams are synchronized. At this boundary, \system coordinates a six-phase distributed state machine over an out-of-band Gloo control plane, replaces the failed communicator, and dynamically rebinds all $L+1$ cached references across FSDP submodules, resuming execution from the committed frontier with zero replay. Any observation falling outside the qualified envelope---such as in-flight operator failures during forward compute or gradient reduction---strictly triggers a fail-stop safe-rejection contract, cleanly delegating to supervisor checkpoint restart to guarantee numerical correctness.

This paper makes four main contributions:
\begin{enumerate}
\item \emph{Framework Dependency Diagnosis \& Non-Invasive Rebinding}: We uncover the dependency caching problem in PyTorch FSDP hybrid sharding, where $L+1$ internal process-group references prevent communicator replacement. We design a dynamic rebinding adapter that updates these references in-situ without modifying user training scripts or breaking compiler symbolic graph capture (\texttt{torch.compile}).
\item \emph{Distributed Agreement Protocol \& Admission Control}: We design a six-phase distributed recovery state machine with out-of-band Gloo coordination, monotonic generation fencing, and fail-stop safe-rejection semantics that guarantees cohort-wide agreement before release, backed by an analytical break-even model.
\item \emph{Empirical Characterization of Checkpoint Storage Backpressure}: In a dedicated microbenchmark on a 1.216B model, we demonstrate that per-update asynchronous checkpointing incurs up to a \DcpAsyncOverhead{} latency overhead, 32.5\,GiB host memory pressure, and \DcpTrafficPerHour{}\,TB/hour of write traffic, refuting the assumption that high-frequency checkpointing can replace in-memory state preservation.
\item \emph{Cluster Evaluation on Mistral-7B}: Across 16 NVIDIA RTX 5880 GPUs training full-parameter Mistral-7B, \system improves goodput by \MacroColdSpeedup$\times$ over cold restart and \MacroNvrxSpeedup$\times$ over NVRx checkpoint restoration at checkpoint age 5, rising to \MacroAgeMaxSpeedup$\times$ at age 18. All ranks maintain bit-identical parameter state across ten successive repairs with zero numerical drift.
\end{enumerate}

\section{Background and Motivation}
\label{sec:background}

\subsection{Distributed Sharded Training and FSDP}
\label{sec:background:fsdp}

To train models exceeding single-GPU memory capacity, ZeRO~\cite{zero2020} and FSDP~\cite{fsdp2023} shard model parameters, gradients, and optimizer states across worker devices. Under Hybrid Sharded Data Parallel (HSDP), GPUs within a physical node form an intra-node sharding group, while corresponding local ranks across different nodes form an inter-node replication group. During forward execution, FSDP all-gathers parameter shards within the node. In backward execution, gradients are reduced and scattered locally, followed by an inter-node all-reduce across replication groups to synchronize parameter updates.

When an inter-node replication collective fails, repairing the communication layer requires replacing the failed communicator while keeping model parameters and optimizer states intact in device memory. However, FSDP module wrappers cache the replication group in internal attributes (such as \texttt{\_inter\_node\_pg}). For a model with $L$ transformer layers and a root wrapper, exactly $L+1$ cached references must be redirected on each affected rank. Omitting this rebinding causes subsequent backward passes to dispatch collectives to the retired communicator handle, crashing with \texttt{DistBackendError: NCCL communicator was aborted}.

\subsection{The Storage Backpressure of Frequent Checkpointing}
\label{sec:background:dcp_pressure}

A common alternative proposal is to checkpoint frequently (e.g., every update) using asynchronous Distributed Checkpointing (DCP) to minimize replay distance. To evaluate the feasibility of this strategy, we conducted a microbenchmark training a 1,216,448,512-parameter causal decoder across four NVIDIA RTX 5880 GPUs using PyTorch 2.11.0 and native DCP with local NVMe storage (Table~\ref{tab:dcp_pressure}).

\begin{table}[t]
\centering
\caption{DCP checkpointing overhead on 1.216B decoder across four GPUs (per-update checkpointing, local NVMe).}
\label{tab:dcp_pressure}
\renewcommand{\arraystretch}{1.08}
\setlength{\tabcolsep}{3pt}
\begin{tabular*}{\columnwidth}{@{\extracolsep{\fill}}lrrrrr@{}}
\toprule
\textbf{Configuration} & \textbf{s/step} & \textbf{Overhead} & \textbf{Fwd/Bwd} & \textbf{PSS} & \textbf{TB/h}\\
\midrule
\DcpPressureRows
\bottomrule
\end{tabular*}
\end{table}

As shown in Table~\ref{tab:dcp_pressure}, baseline training alone requires 1.024\,s per update. Synchronous checkpointing increases step latency to 9.960\,s (+870.8\%). Standard asynchronous staging via background threads or processes exacerbates the stall (+1043.5\% and +1090.2\%) due to thread scheduling contention and host memory copying overhead (+240\% forward/backward increase). Even native cached-process staging incurs a +656.7\% latency overhead (7.747\,s/step), increases forward/backward compute by 17.96\%, elevates host Proportional Set Size (PSS) from 9.07 to 32.52\,GiB, and generates 3.39\,TB/hour of write traffic. Furthermore, completed snapshots consistently lag execution by two full updates, meaning failures still require replaying two updates.

Extrapolating this to our 16-GPU Mistral-7B workload, saving per-rank state (10.9\,GB, or 43.6\,GB/node) over local NVMe ($1.70\,\text{GB/s}$ write bandwidth) requires $\approx 25.6\,\text{s}$ per update, imposing an $+11.6\%$ continuous I/O penalty. These empirical results prove that frequent checkpointing cannot replace in-memory state preservation on high-throughput workloads.

\begin{figure}[t]
\centering
\includegraphics[width=\columnwidth]{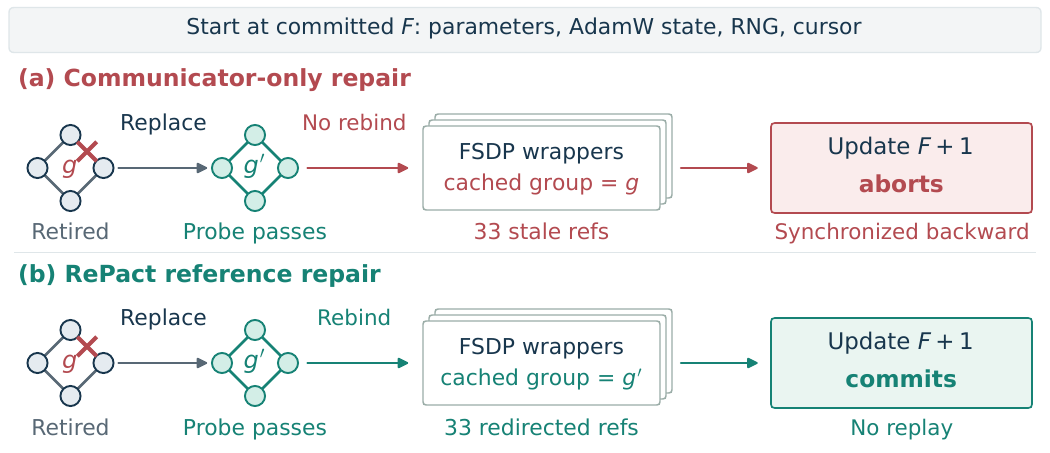}
\caption{Contrast between communicator-only replacement and \system reference rebinding. Replacing the communicator alone passes isolated group probes, but crashes in synchronized backward on 33 stale FSDP references. \system dynamically redirects these references to resume training from frontier $F$ with zero replay.}
\label{fig:architecture}
\end{figure}

\subsection{Cross-Stack Driver Watchdog Divergence}
\label{sec:background:cross_stack}

State recoverability also depends on accelerator driver behavior. We evaluated in-process communicator destruction and recreation across two hardware stacks: NVIDIA A100 (CUDA 12.1 / NCCL 2.21.5) and Huawei Ascend 910B (CANN 9.1.0 / HCCL 9.1.0) under identical two-rank DDP setups.

After an incomplete collective timeout, calling \texttt{destroy\_process\_group()} followed by \texttt{init\_process\_group()} succeeds cleanly on NVIDIA A100 across five fresh pairs, allowing workers to execute 128 subsequent collective epochs without replay. In contrast, on Ascend 910B, driver task queues retain pending work items from the aborted collective, triggering a hardware watchdog timeout that kills the process before recreation finishes. However, on recurrent computational graph workloads, this relationship inverts: Ascend's driver admits graph re-execution while A100 requires a clean process restart. This proves that recovery qualification is not a static hardware attribute, but an empirical contract between the execution context and the recovery action.

\begin{table}[t]
\centering
\caption{Cross-stack recovery policy matrix across accelerator platforms (4 DDP trials per cell; $L_1$: in-memory preserve, $L_2$: checkpoint restart).}
\label{tab:policy_matrix}
\renewcommand{\arraystretch}{1.08}
\setlength{\tabcolsep}{4pt}
\begin{tabular*}{\columnwidth}{@{\extracolsep{\fill}}lllrr@{}}
\toprule
\textbf{Stack} & \textbf{Policy} & \textbf{Action} & \textbf{Valid} & \textbf{Replayed Updates}\\
\midrule
\PolicyMatrixRows
\bottomrule
\end{tabular*}
\end{table}

\section{\system Architecture}
\label{sec:design}

\subsection{State Model and Qualification Contract}
\label{sec:design:model}

\system models distributed training state as a composition of computational state $\mathcal{S}_{\mathrm{app}}$ and runtime resources $\mathcal{S}_{\mathrm{runtime}}$:
\begin{equation}
\mathcal{S} = \langle \mathcal{S}_{\mathrm{app}}, \mathcal{S}_{\mathrm{runtime}} \rangle,
\end{equation}
\begin{equation}
\mathcal{S}_{\mathrm{app}} = \{ W, M, V, \mathcal{S}_{\mathrm{RNG}}, \kappa, F, \tau_{\mathrm{opt}} \},
\end{equation}
\begin{equation}
\mathcal{S}_{\mathrm{runtime}} = \{ \mathcal{G}_{\mathrm{comm}}, \Sigma_{\mathrm{stream}}, \mathcal{R}_{\mathrm{ref}} \}.
\end{equation}
Here, $W$ denotes sharded model weights, $M$ and $V$ are AdamW first and second moments, $\mathcal{S}_{\mathrm{RNG}}$ is device and CPU random state, $\kappa$ is the dataset token cursor, $F$ is the committed optimizer frontier index, and $\tau_{\mathrm{opt}}$ is the optimizer step counter. Runtime resources include the active communicator $\mathcal{G}_{\mathrm{comm}}$, CUDA streams $\Sigma_{\mathrm{stream}}$, and framework references $\mathcal{R}_{\mathrm{ref}}$.

Recovery decisions are mediated by an observation key $\mathcal{K} = \langle p, r, \phi, o \rangle$, where $p$ is the platform stack, $r$ is the resource profile, $\phi$ is the failure phase, and $o$ is the continuation objective. An envelope entry associates $\mathcal{K}$ with qualified recovery actions:
\begin{itemize}
\item \textbf{Preserve ($L_1$)}: Applicable when failure occurs at a committed optimizer boundary ($\phi = \text{committed}$). Workers retain $\mathcal{S}_{\mathrm{app}}$ in device memory, destroy and rebuild $\mathcal{G}_{\mathrm{comm}}$, rebind $\mathcal{R}_{\mathrm{ref}}$, and resume from frontier $F$ with zero replay.
\item \textbf{Restart ($L_2$)}: Triggered when failure occurs during uncommitted compute ($\phi \in \{ \text{forward}, \text{backward} \}$). Workers strictly issue a fail-stop rejection, delegating to the cluster supervisor to reload checkpoint $C$ and replay $A = F - C$ updates.
\end{itemize}

\subsection{Break-Even Sensitivity Analysis}
\label{sec:design:breakeven}

To quantify the operational value of in-memory recovery, we formulate the net cluster time saved $\Delta T$ over a training campaign of baseline duration $T_0$:
\begin{equation}
\label{eq:breakeven}
\Delta T = N_{\mathrm{eligible}} \cdot S - N_{\mathrm{other}} \cdot C - h \cdot T_0,
\end{equation}
where $N_{\mathrm{eligible}}$ is the number of communication failures occurring at committed boundaries; $S = A \cdot t_{\mathrm{update}} - T_{\mathrm{repair}}$ is the time saved per eligible recovery event; $N_{\mathrm{other}}$ is the number of uncommitted failures; $C = T_{\mathrm{detect}} + T_{\mathrm{reject}}$ is the safe rejection overhead ($C < 0.1$\,s); and $h$ is the fractional runtime overhead of boundary guards.

For a 24-hour training run ($T_0 = 86,400$\,s), setting $h = 0.002$ ($0.2\%$, matching our measured 16-GPU guard overhead of $+0.200\%$ and $-0.412\%$) incurs a daily monitoring cost of $h \cdot T_0 \approx 172.8$\,s. With checkpoint age $A=5$ and update duration $t_{\mathrm{update}} = 220$\,s, each in-memory recovery saves $S = 5 \times 220 - 0.77 \approx 1,099.2$\,s. Thus, just \textbf{one eligible failure every 6.3 days} ($1,099.2 / 172.8$) offsets continuous guard monitoring across the cluster.

Grounded in Meta's Llama 3 405B training logs~\cite{llama3herd2024} (7.76 interruptions/day, with network and NCCL watchdog issues accounting for 0.916 events/day): if $2\%$ of interruptions coincide with committed boundaries ($0.155\,\text{events/day}$, or one incident every 6.45 days), daily replay savings ($0.155 \times 1,099.2 = 170.4\,\text{s/day}$) approximately offsets ($98.6\%$) continuous monitoring cost ($172.8\,\text{s/day}$). If $10\%$ of failures coincide with boundary collectives ($0.776\,\text{events/day}$), \system achieves a gross saving of $852.9\,\text{s/day}$ and a net saving of $680.2\,\text{s/day}$ ($\approx 3.9\times$ daily monitoring cost).

\subsection{Distributed Recovery Protocol}
\label{sec:design:protocol}

Recovery is governed by a six-phase distributed state machine across the full cluster $W$ ($|W|=16$) and the affected replication subgroup $R \subset W$ ($|R|=4$):
\begin{enumerate}
\item \textbf{Observe}: Workers detect an incomplete collective timeout at a committed boundary. Under \texttt{TORCH\_NCCL\_BLOCKING\_WAIT=1}, PyTorch invokes \texttt{ncclCommAbort()} and raises an exception in the main thread.
\item \textbf{Guard Check \& Stream Sync}: Workers execute \texttt{torch.cuda.synchronize()} to flush device queues, verifying all FSDP modules are \texttt{IDLE} with zero outstanding gradients.
\item \textbf{Control Barrier \& Frontier Agreement}: All ranks in $W$ synchronize over an out-of-band Gloo TCP channel. Rank 0 broadcasts committed frontier index $F$ and generation counter $g$. Ranks assert exact agreement on $F$; any mismatch aborts immediately.
\item \textbf{Communicator Replacement}: Ranks in $R$ destroy the retired communicator and initialize a replacement with identical rank membership over the control plane.
\item \textbf{Reference Rebinding}: Ranks in $R$ traverse the local FSDP module hierarchy, updating all $L+1$ cached process-group references to point to the replacement communicator.
\item \textbf{Validation Barrier \& Release}: Ranks in $R$ execute an integer all-reduce on the new communicator to verify hardware readiness. Upon success, all ranks in $W$ synchronize on a Gloo release barrier and resume training.
\end{enumerate}

\paragraph{Safety Invariant} If any worker encounters an exception or timeout during phases 2--6, it terminates immediately. The Gloo control channel detects worker loss, causing all surviving peers to exit within bounded deadlines, cleanly delegating to supervisor checkpoint restart without split-brain execution.

\subsection{The FSDP Reference Rebinding Invariant}
\label{sec:design:invariant}

Let $\mathcal{O}_{\mathrm{FSDP}}$ be the set of FSDP module wrappers, parameter handles, and execution metadata. Rebinding enforces the invariant:
\begin{equation}
\begin{aligned}
\forall r \in \mathcal{O}_{\mathrm{FSDP}}, \quad &(r = g_{\mathrm{old}} \implies r \leftarrow g_{\mathrm{new}}) \\
&\land (r \ne g_{\mathrm{old}} \implies r \text{ unchanged}).
\end{aligned}
\end{equation}
On return, every inspected reference formerly naming $g_{\mathrm{old}}$ must name $g_{\mathrm{new}}$, and no reference to $g_{\mathrm{old}}$ may remain. In our 32-layer Mistral-7B setup, exactly 33 internal references (one per layer wrapper plus root) are dynamically redirected.

\section{Evaluation}
\label{sec:eval}

We evaluate \system on real hardware across eight core dimensions: (1) 16-GPU full-parameter Mistral-7B training goodput; (2) checkpoint-age sensitivity; (3) recovery breakdown scaling across topologies; (4) causal necessity of reference rebinding; (5) safe rejection of in-flight faults; (6) repeated lifecycle stability and numerical drift; (7) fault-free guard overhead; and (8) architectural trade-offs versus wrapper indirection (\texttt{torchft}).

\subsection{Experimental Setup}
\label{sec:eval:setup}

\paragraph{Testbeds} The primary 16-GPU testbed comprises four nodes, each equipped with four NVIDIA RTX 5880 Ada GPUs (48\,GiB VRAM, driver 570.169), interconnected via 1\,Gbps Ethernet (PyTorch 2.5.1, CUDA 12.1, NCCL 2.21.5). FSDP hybrid sharding uses 4-way intra-node sharding and 4-way inter-node replication. Cross-stack evaluations use a four-GPU NVIDIA A100 server and an eight-NPU Huawei Ascend 910B server.

\paragraph{Workloads} The macro workload trains all \MacroParameters{} parameters of Mistral-7B-v0.3~\cite{mistral2023,mistralweights} on WikiText-103~\cite{wikitext} using sequence length 512, microbatch 1, and 8 accumulation steps (65,536 tokens/update). Model parameters and AdamW moments use BF16 with activation checkpointing. An optimizer update takes $\approx 220$\,s.

\subsection{Full-Model Training on 16 GPUs}
\label{sec:eval:macro}

\begin{figure}[t]
\centering
\includegraphics[width=\columnwidth]{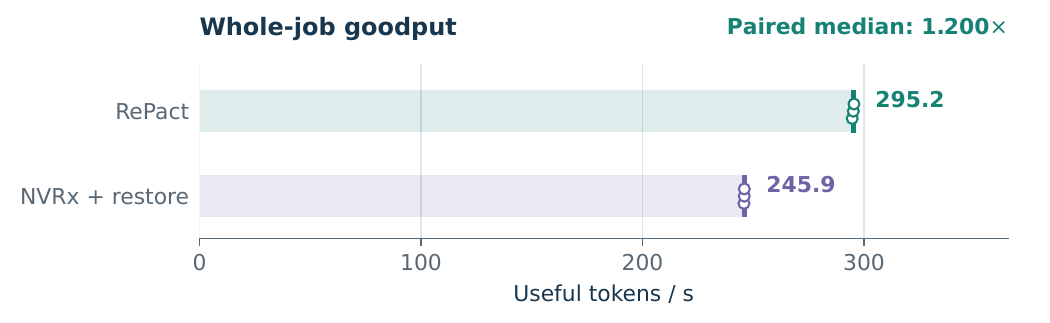}
\caption{Full-model goodput on 16 GPUs training Mistral-7B. (a) Whole-run token throughput; (b) Paired whole-run speedup ratios over cold restart and NVRx.}
\label{fig:goodput}
\end{figure}

Table~\ref{tab:goodput25} and Figure~\ref{fig:goodput} present whole-run results over 25 optimizer updates (1,638,400 tokens), with a checkpoint written at update 10 and an inter-node collective fault injected after update 15 ($A=5$).

\begin{table}[t]
\centering
\caption{Whole-run training goodput on 16 GPUs (Mistral-7B, 25 updates, checkpoint at update 10, fault after update 15, $A=5$).}
\label{tab:goodput25}
\renewcommand{\arraystretch}{1.08}
\setlength{\tabcolsep}{3pt}
\begin{tabular*}{\columnwidth}{@{\extracolsep{\fill}}lrrrrrr@{}}
\toprule
& \multicolumn{3}{c}{\textbf{Whole Run (s)}} & \multicolumn{3}{c}{\textbf{Fault-to-Frontier (s)}}\\
\cmidrule(lr){2-4} \cmidrule(lr){5-7}
\textbf{Block} & \textbf{\system} & \textbf{Cold} & \textbf{NVRx} & \textbf{\system} & \textbf{Cold} & \textbf{NVRx}\\
\midrule
\MacroResultsRows
\bottomrule
\end{tabular*}
\end{table}

\system achieves a paired median whole-run goodput speedup of \textbf{\MacroColdSpeedup$\times$} over cold restart (range \MacroColdRange$\times$, saving \MacroColdSavedMinutes{} minutes) and \textbf{\MacroNvrxSpeedup$\times$} over NVRx (range \MacroNvrxRange$\times$, saving \MacroNvrxSavedMinutes{} minutes). Looking at the fault-to-frontier duration, \system completes the window in $\approx 2,367$\,s, compared to $\approx 3,500$\,s for cold restart and $\approx 3,466$\,s for NVRx ($1.47\times$ faster). While cold restart and NVRx must reload checkpoint 10 and replay five 220-second updates, \system's sub-second repair (0.672--0.930\,s across blocks) eliminates 1,100 seconds of redundant computation, immediately advancing to update 16.

\subsection{Checkpoint-Age Sensitivity}
\label{sec:eval:macro_age}

\begin{figure*}[t]
\centering
\includegraphics[width=0.81\textwidth]{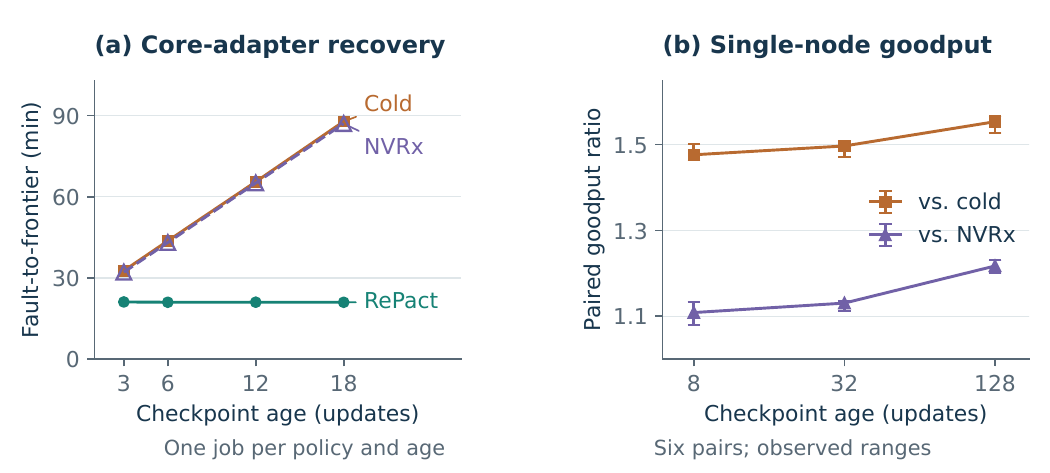}
\caption{Checkpoint-age sensitivity on 16 GPUs. \system's fault-to-frontier duration remains flat ($\approx$21 minutes), while restart paths scale linearly with checkpoint age $A$.}
\label{fig:age_divergence}
\end{figure*}

To evaluate scaling with checkpoint age, we conducted a 12-job sweep across ages $A \in \{3, 6, 12, 18\}$ (Table~\ref{tab:macro_age} and Figure~\ref{fig:age_divergence}).

\begin{table}[t]
\centering
\caption{Checkpoint-age sensitivity on 16 GPUs (Mistral-7B, fault after update 20, target frontier at update 25).}
\label{tab:macro_age}
\renewcommand{\arraystretch}{1.08}
\setlength{\tabcolsep}{3pt}
\begin{tabular*}{\columnwidth}{@{\extracolsep{\fill}}lrrrrrr@{}}
\toprule
& \multicolumn{3}{c}{\textbf{Whole Run (s)}} & \multicolumn{3}{c}{\textbf{Fault-to-Frontier (s)}}\\
\cmidrule(lr){2-4} \cmidrule(lr){5-7}
\textbf{Age $A$} & \textbf{\system} & \textbf{Cold} & \textbf{NVRx} & \textbf{\system} & \textbf{Cold} & \textbf{NVRx}\\
\midrule
\MacroAgeTableRows
\bottomrule
\end{tabular*}
\end{table}

Because \system eliminates replay, its fault-to-frontier duration remains constant at \textbf{\MacroAgeMinCompletionMinutes{}--\MacroAgeMaxCompletionMinutes{} minutes} across all ages. In contrast, cold restart scales linearly from 32.65 minutes at $A=3$ to 87.81 minutes at $A=18$. At $A=18$, \system completes the recovery window \textbf{\MacroAgeMaxCompletionSpeedup$\times$ faster} than cold restart (saving 66.82 minutes per failure), delivering a whole-run speedup of \textbf{\MacroAgeMaxSpeedup$\times$}.

\subsection{Recovery Topology Scaling}
\label{sec:eval:topology}

\begin{figure}[t]
\centering
\includegraphics[width=\columnwidth]{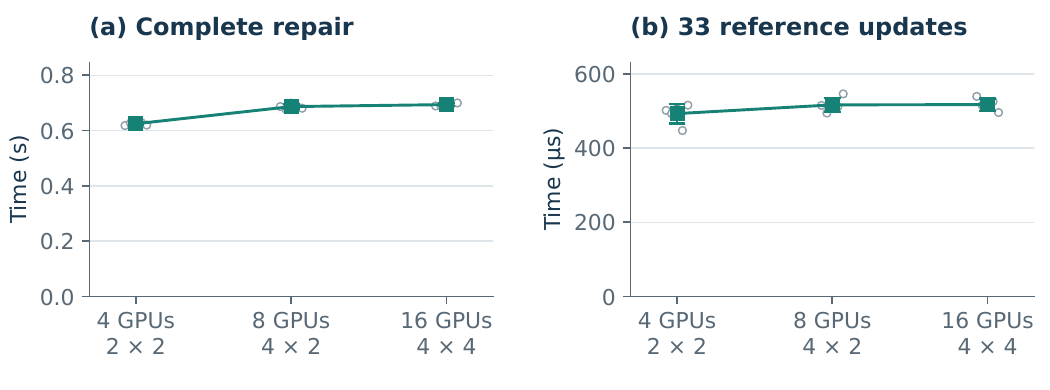}
\caption{Recovery-stage scaling across topologies (4, 8, and 16 GPUs). Reference rebinding remains invariant at $\approx$0.5\,ms, with sub-second total repair.}
\label{fig:topology_scaling}
\end{figure}

To evaluate how \system scales with cluster topology, we measured breakdown timings on a 32-layer causal decoder across 4 GPUs (2 nodes $\times$ 2 GPUs), 8 GPUs (4 nodes $\times$ 2 GPUs), and 16 GPUs (4 nodes $\times$ 4 GPUs), conducting five trials per layout (Table~\ref{tab:topology_scaling} and Figure~\ref{fig:topology_scaling}).

\begin{table}[t]
\centering
\caption{Recovery-stage breakdown on small decoder across GPU topologies (slowest rank values in milliseconds, mean over 5 trials).}
\label{tab:topology_scaling}
\renewcommand{\arraystretch}{1.08}
\setlength{\tabcolsep}{2pt}
\begin{tabular*}{\columnwidth}{@{\extracolsep{\fill}}lrrrrr@{}}
\toprule
\textbf{Scale} & \textbf{Admit} & \textbf{Retire} & \textbf{NCCL} & \textbf{Rebind} & \textbf{Total}\\
\midrule
\TopoScalingRows
\bottomrule
\end{tabular*}
\end{table}

Gloo admission scales from 9.36\,ms (4 GPUs) to 26.82\,ms (16 GPUs). Communicator retirement takes 531.72--542.60\,ms to cleanly release driver resources. NCCL rebuild requires 78.98\,ms (2-member) to 118.57--124.18\,ms (4-member). Crucially, 33-reference rebinding remains strictly constant at \textbf{0.493--0.518\,ms} across all scales ($493\text{--}518\,\mu\text{s}$), reflecting constant CPU-local pointer inspection at fixed 33-reference depth. Total repair remains sub-second (0.625--0.694\,s).

\subsection{Why FSDP References Must Be Rebound}
\label{sec:eval:ablation}

To isolate the necessity of reference rebinding, we executed an ablation where group replacement succeeded and passed its integer collective check, but \texttt{rebind\_fsdp\_states()} was skipped. During update 9, the first seven microbatches completed under \texttt{no\_sync}. At microbatch 8, synchronized backward propagation invoked inter-node gradient reduction. Because the 33 internal references still named the retired communicator, execution crashed immediately with \texttt{DistBackendError: NCCL communicator was aborted}. This causal ablation confirms that communicator reconstruction alone is insufficient; dynamic framework rebinding is mandatory for continuation.

\subsection{Operator-Level Faults and Safe Rejection}
\label{sec:eval:operator_faults}

We evaluated \system under uncommitted in-flight operator failures on a 4-layer causal decoder across four GPUs:

\noindent\textbf{In-Flight Collective Rejection:} We injected communication timeouts during forward \texttt{all-gather} and backward \texttt{reduce-scatter}. Boundary guards detected active module state (\texttt{FORWARD} / \texttt{BACKWARD}) with non-zero unreduced gradients, strictly issuing \texttt{reject\_uncommitted\_boundary}. Workers terminated with non-zero exit codes, cleanly delegating to supervisor checkpoint reload from step 1 to 5, with all 8/8 rank state checks passing bitwise.

\noindent\textbf{Post-Commit Metric Reduction:} When a collective timeout was injected during post-commit loss logging across replication shards (with modules \texttt{IDLE} and gradients cleared), \system admitted $L_1$ repair, rebuilt the communicator, rebound references, and resumed to step 5 without replay, passing all 4/4 state checks.

\subsection{Repeated Lifecycle and Drift Verification}
\label{sec:eval:lifecycle}

On 16 GPUs training Mistral-7B, we injected ten successive collective faults across 25 updates. All ten in-process repairs completed successfully (\MacroRepeatedMinRepair--\MacroRepeatedMaxRepair\,s per repair). GPU allocated memory remained rock-solid at 10.188\,GiB, reserved memory remained at 30.887\,GiB (Table~\ref{tab:lifecycle}), and all 16 ranks matched the uninterrupted reference at the final frontier with zero numerical difference.

\begin{table}[t]
\centering
\caption{Quiescent resources across ten repairs (Mistral-7B, 16 GPUs).}
\label{tab:lifecycle}
\renewcommand{\arraystretch}{1.08}
\setlength{\tabcolsep}{4pt}
\begin{tabular*}{\columnwidth}{@{\extracolsep{\fill}}lrr@{}}
\toprule
\textbf{Resource} & \textbf{Before Faults} & \textbf{After 10 Repairs}\\
\midrule
\MacroLifecycleRows
\bottomrule
\end{tabular*}
\end{table}

In an automatic-precision drift study on a 4-GPU 32-layer decoder (36 jobs, 12 matched triplets across 3 seeds), all 96 final rank tensor comparisons matched the uninterrupted reference bitwise ($L_2 = 0$).

\subsection{Fault-Free Guard Overhead}
\label{sec:eval:overhead}

\begin{table}[t]
\centering
\caption{Single-node fast-guard overhead across hardware stacks.}
\label{tab:guard_overhead}
\renewcommand{\arraystretch}{1.08}
\setlength{\tabcolsep}{4pt}
\begin{tabular*}{\columnwidth}{@{}lrrl@{}}
\toprule
\textbf{Cell} & \textbf{Pairs} & \textbf{Median} & \textbf{95\% Interval}\\
\midrule
\SingleGuardRows
\bottomrule
\end{tabular*}
\end{table}

As reported in Table~\ref{tab:guard_overhead}, observed guard overhead medians range from $-1.339\%$ to $1.895\%$ on single-node cells. On the 16-GPU workload, two native/guarded pairs of 60 updates showed measured paired effects of $+0.200\%$ and $-0.412\%$, confirming that boundary checks impose negligible overhead on production steps.

\subsection{Architectural Comparison with \texttt{torchft}}
\label{sec:eval:torchft}

\begin{table}[t]
\centering
\caption{Architectural and engineering trade-offs between \system and upstream \texttt{torchft}.}
\label{tab:torchft_comparison}
\renewcommand{\arraystretch}{1.08}
\setlength{\tabcolsep}{3pt}
\begin{tabularx}{\columnwidth}{@{}lXX@{}}
\toprule
\textbf{Dimension / Feature} & \system & \texttt{torchft} \\
\midrule
\TorchftComparisonRows
\bottomrule
\end{tabularx}
\end{table}

We contextualize \system against upstream \texttt{torchft}~\cite{torchft2024} (Meta PyTorch), which targets in-process recovery via communication wrapper indirection rather than dynamic attribute rebinding (Table~\ref{tab:torchft_comparison}).

In an evaluation on 4 GPUs under PyTorch 2.11.0 + NCCL 2.28.9 with a 32-layer causal decoder, \texttt{torchft}'s \texttt{ProcessGroupWrapper} demonstrated successful reconfiguration in 3/3 trials ($T_{\mathrm{repair}} \approx 0.71$\,s; resumed in 12.9\,s; $L_2 = 0$). In contrast, \system's native hook prototype encountered pre-repair Gloo timeouts under PyTorch 2.11's exception handling, failing to enter repair dispatch (0/3). This highlights a fundamental systems trade-off:

\noindent(1)~\emph{Decoupling vs. Zero-Code Integration}: \texttt{torchft}'s wrapper indirection isolates application control flow from transport aborts on newer stacks, but requires intrusive changes ($\sim$15--25 lines wrapping \texttt{init\_process\_group} and FSDP handles). Conversely, \system attaches to stock FSDP via hooks with \textbf{zero code changes}, though its exception capture couples to runtime semantics (validated across our 16-GPU PyTorch 2.5.1 campaigns).

\noindent(2)~\emph{Compiler Tracing}: Under \texttt{torch.compile(mode="reduce-overhead")}, custom wrappers require functional collective registration to avoid TorchDynamo graph breaks. In contrast, \system preserves native \texttt{ProcessGroupNCCL} C++ instances, enabling Dynamo to capture whole-graph collectives with \textbf{zero wrapper-induced graph breaks}.

\section{Related Work}
\label{sec:related}

\noindent\textbf{HPC Resilience \& Communicator Recovery:} User-Level Fault Mitigation (ULFM)~\cite{ulfm2013} extended MPI with communicator revocation and shrink primitives to enable continuation without restart. Reinit++~\cite{reinit2020} evaluated in-memory restart to eliminate redeployment. \system builds on this transport/state separation, resolving the deep learning challenge where frameworks cache communication handles within sharded model states.

\noindent\textbf{High-Performance Checkpointing:} Scalable checkpointing systems like SCR~\cite{scr2010} and VeloC~\cite{veloc2019} leverage hierarchical storage to mitigate I/O overhead. In deep learning, CheckFreq~\cite{checkfreq2021} pipelines snapshotting with compute, and Gemini~\cite{gemini2023} caches checkpoints in host memory. As shown in Section~\ref{sec:background:dcp_pressure}, even asynchronous snapshots incur compute stalls and multi-terabyte traffic at high frequencies. \system complements checkpointing with a zero-I/O fast path for boundary faults.

\noindent\textbf{Distributed Deep Learning Fault Tolerance:} Bamboo~\cite{bamboo2023} uses redundant compute; Oobleck~\cite{oobleck2023} reconfigures pipeline templates; Chameleon~\cite{chameleon2026} searches execution plans; ReCoVer~\cite{recover2026} explores trajectory preservation under HSDP; \texttt{torchft}~\cite{torchft2024} wraps process groups in indirection layers; NVRx~\cite{nvrx2024} coordinates restart with checkpoint reload. \system uniquely formalizes context qualification and resolves FSDP references without wrapper encapsulation.

\section{Limitations and Discussion}
\label{sec:limitations}

While \system delivers substantial goodput gains, we explicitly identify its boundary limitations:
(1)~\emph{Envelope Scope}: \system targets collective failures at committed boundaries; in-flight failures during uncommitted compute safely reject to supervisor checkpoint restart. Mid-step discard is left to future work.
(2)~\emph{Framework Specificity}: Direct reference rebinding inspects internal FSDP1 attributes (\texttt{\_inter\_node\_pg}, \texttt{FlatParamHandle}). While verified across PyTorch 2.1--2.11, future architectures (e.g., FSDP2/DTensor) require corresponding adapter updates.
(3)~\emph{Failure Scope}: \system addresses transient timeouts, hangs, and partitions; unrecoverable node crashes and GPU hardware ECC faults require process relaunch and checkpoint restoration.
(4)~\emph{Interconnect Context}: While evaluated over 1\,Gbps Ethernet, \system's advantage stems from eliminating $A$ replayed steps, providing proportional savings on multi-gigabit fabrics.

\section{Conclusion}
\label{sec:conclusion}

Transient communication failures do not corrupt committed training progress in device memory. \system operationalizes this insight for sharded training by coupling context qualification with dynamic reference rebinding. By repairing $L+1$ cached FSDP references in surviving processes, \system eliminates checkpoint replay, delivering up to \MacroAgeMaxSpeedup$\times$ goodput improvement on 16 GPUs with zero numerical drift and sub-second recovery across cluster topologies. In microbenchmarks, \system demonstrates that frequent checkpointing incurs severe backpressure on short-step workloads, establishing in-memory state preservation as an essential complement to periodic checkpointing. These results validate framework-aware state preservation as a practical foundation for resilient distributed deep learning.

\section*{Acknowledgment}
OpenAI ChatGPT was used during the preparation of this manuscript to assist with language polishing and clarity of expression. The authors reviewed and verified all AI-assisted revisions and take full responsibility for the final content.

\bibliographystyle{IEEEtran}
\bibliography{references}

\end{document}